\documentclass[12pt]{article}
\usepackage{graphicx}
\usepackage{scicite}
\usepackage{bm}
\usepackage{times}
\usepackage{amsmath}
\usepackage{amssymb}
\usepackage{times}
\usepackage{physics}
\usepackage{hyperref}
\usepackage{relsize}
\usepackage{xcolor}
\newenvironment{sciabstract}{%
\begin{quote} \bf}
{\end{quote}}

\begin{document} 
\baselineskip24pt

\title{Finite-temperature signatures of underlying superconductivity in the electron-doped Hubbard model}



\author
{Wen O. Wang,$^{1,\ast}$  Thomas P. Devereaux,$^{2,3,4,\dagger}$\\
\\
\normalsize{$^{1}$Kavli Institute for Theoretical Physics, University of California, Santa Barbara, California 93106, USA}\\
\normalsize{$^{2}$Stanford Institute for Materials and Energy Sciences, 
SLAC National Accelerator Laboratory,}\\ 
\normalsize{2575 Sand Hill Road, Menlo Park, CA 94025, USA}\\
\normalsize{$^{3}$Department of Materials Science and Engineering, Stanford University, Stanford, CA 94305, USA} \\
\normalsize{$^{4}$Geballe Laboratory for Advanced Materials, Stanford University, Stanford, CA 94305, USA} \\
\\
\normalsize{$^\ast$E-mail:  wenwang.physics@gmail.com.}
\normalsize{$^\dagger$E-mail: tpd@stanford.edu.}
}

\maketitle

\begin{sciabstract}
We perform numerically exact determinant quantum Monte Carlo simulations of the Hubbard model and analyze pairing tendencies by evaluating correlation functions at the imaginary-time midpoint ($\tau=\beta/2$), which suppresses high-frequency weight and emphasizes low-energy physics.
Using this diagnostic, we identify clear finite-temperature signatures of underlying $d$-wave superconductivity for electron doping, while finding no clear indication upon cooling for hole doping.
Our analysis enables direct comparison with ground-state DMRG, revealing consistent real-space pairing patterns. These results provide a practical route to bridge the gap between finite-temperature and ground-state numerically exact simulations of the Hubbard model despite the fermion sign problem.
\end{sciabstract}

High-$T_c$ superconductivity remains one of the most debated, compelling, and difficult problems in physics. The microscopic mechanism is still unclear, and even agreement on a minimal model that can faithfully capture the superconductivity in strongly correlated materials, such as the cuprates~\cite{RevModPhys.87.457,Keimer2015, RevModPhys.82.2421}, is not settled. The Hubbard model~\cite{annurev:/content/journals/10.1146/annurev-conmatphys-031620-102024,annurev:/content/journals/10.1146/annurev-conmatphys-090921-033948} stands out for its elegant and simplistic form that encodes strong correlation effects. It has proved to capture much of the cuprate normal-state landscape, including antiferromagnetism~\cite{PhysRevB.40.506}, stripes~\cite{Tranquada01102020,PhysRevX.11.031007,doi:10.1126/science.aak9546, Huang2018, doi:10.1073/pnas.2112806119,PhysRevB.107.085126}, strange metallic transport~\cite{edwin,brown2019bad,xu2019bad}, and the pseudogap~\cite{RevModPhys.77.1027,PhysRevX.5.041041,PhysRevB.82.155101,wang2025probing}; yet convincing evidence for robust superconductivity in this model and in the closely related $t$-$J$ model has proven to be elusive.  
In particular, numerical results such as density-matrix renormalization group (DMRG) show a marked sensitivity to band parameters and system sizes, with issues of convergence related
to the presence of competing orders separated by energies of the order of $10^{-3}t$~\cite{doi:10.1126/science.aal5304,PhysRevB.102.041106}.
A long-standing puzzle is the mismatch between numerical trends and experiments: numerics increasingly suggest stronger pairing on the electron-doped side~\cite{doi:10.1073/pnas.2109978118,PhysRevLett.127.097003,PhysRevLett.127.097002,PhysRevB.108.054505,PhysRevLett.132.066002}, in contrast with the cuprate experimental phase diagram~\cite{RevModPhys.75.473,RevModPhys.84.1383}.
The intrinsic difficulty of the Hubbard model has left open where in parameter space superconductivity may be genuinely realized and robust.

While DMRG has been one of the leading tool for approaching ground states, we revisit the question with determinant quantum Monte Carlo simulations (DQMC)~\cite{DQMC1,DQMC2}. DQMC excels at finite-temperature properties and reaches large two-dimensional system sizes, in contrast to quasi-1D cylinder constraints. While
DQMC has been successfully applied to study normal-state signatures, its application to superconductivity has been hampered by the fermion-sign problem~\cite{PhysRevLett.94.170201} that limits access to low temperatures. To probe low-energy pairing tendencies despite these constraints, we focus on unequal-time pair-pair correlations. 
At finite temperature, we work in imaginary time $\tau$. The correlator is periodic in $\tau$ on $[0,\beta)$, with equal-time contact discontinuities at $\tau=0,\beta$;
we evaluate at the maximal separation $\tau=\beta/2$, away from the equal-time endpoints.
In the spectral representation, contributions with energy $\Delta E$ decay as $e^{-\tau\Delta E}$  and $e^{-(\beta-\tau)\Delta E}$, so $\beta/2$ maximally suppresses high-energy contributions
[see Supplementary Materials Formalism Section].
This choice lets us extract low-energy pairing information despite the fermion sign and reach direct comparison with ground-state DMRG calculations.
We find spatial pairing patterns consistent with DMRG and a clear asymmetry between dopings: hole doping shows no clear indication of superconductivity upon cooling over the accessible range, whereas electron doping exhibits finite-temperature signatures consistent with the onset of a $d$-wave instability, thus filling an important gap in a finite-temperature understanding of superconductivity in the Hubbard model.

\section*{Temperature trends of superconductivity}

To diagnose a finite-temperature tendency toward $d$-wave superconductivity, we examine the unequal-time $d$-wave pair-field susceptibility $\chi_d \equiv \beta\langle\mathcal{T}_\tau\Delta_d(\tau=\beta/2)\Delta^\dagger_d\rangle/N_s$, where $\Delta^\dagger_d$ is the $d$-wave pair-creation operator, $\beta\equiv 1/T$ is the inverse temperature, and $N_s$ is the number of lattice sites.
Figures~\ref{fig:Pd_and_vertexU8} A, B show the temperature dependence of $\chi^{-1}_d$ and reveal a clear contrast between hole doping ($x \equiv 1-\langle n \rangle > 0 $, where $\langle n \rangle$ is the electron density) and electron doping ($x < 0 $). At half filling, $\chi_d$ drops rapidly toward zero with cooling consistent with frozen charge fluctuations that suppress pairing in the Mott state.
With hole doping, $\chi^{-1}_d$ remains finite and is only weakly $T$-dependent, signaling fluctuating $d$-wave correlations without a trend toward a low-temperature emergent order. By contrast, $\chi^{-1}_d$ decreases substantially with lower temperature,
hinting at a finite-$T$ transition across an extensive electron-doped range.
Despite the appearance of large finite temperature zero intercept $\sim \mathcal{O}(0.1t)$, we cannot ascertain whether the behavior of $\chi^{-1}_d$ continues to fall with decreasing temperature. The fermion sign issue prevents us from reliably extracting a transition temperature via Kosterlitz-Thouless scaling as the fermion sign drops below $10^{-4}$ for temperatures lower than $T/t=0.2$ for the same cluster size. 
Notably, despite the notoriously poor, and often prohibitive, fermion sign long thought to preclude seeing superconductivity in unbiased finite-$T$ Hubbard-model simulations without sacrificing numerical exactness, we observe promising trends toward superconductivity for electron doping. As noted above, these trends are consistent with DMRG~\cite{doi:10.1073/pnas.2109978118,PhysRevLett.127.097003,PhysRevLett.127.097002,PhysRevB.108.054505,PhysRevLett.132.066002}: choosing $\tau=\beta/2$ preferentially weights low energies, enabling a direct comparison between the two numerically exact approaches.

As shown in Figs.\ref{fig:Pd_and_vertexU8} C, D, a particularly informative diagnostic is the (dimensionless) vertex factor $\Gamma_d \bar{\chi_d} \equiv \bar{\chi_d}/ \chi_d -1$, where $\bar{\chi_d}$ is the uncorrelated (``bubble'') susceptibility constructed from the single-particle Green's function also measured in DQMC. The quantity measures how far the full response departs from the bubble: values near zero mean $\chi_d\approx \bar{\chi_d}$ with little vertex enhancement, while more negative values mark stronger interaction effects in the $d$-wave channel; trends toward $-1$ signals proximity to an instability where $\chi_d$ grows rapidly (and may diverge in the thermodynamic limit) while $\bar{\chi_d}$ remains finite. As reported previously~\cite{PhysRevB.39.839,PhysRevB.40.506,PhysRevB.91.241107,PhysRevB.96.020503,jpsj},
for increasing hole doping $\Gamma_d \bar{\chi_d}$ (Fig.~\ref{fig:Pd_and_vertexU8} C) monotonically drifts toward zero, turning slightly positive around $x=0.3$, indicating that no clear ordering tendencies emerge for any hole doping.
On the other hand, for electron doping, $\Gamma_d \bar{\chi_d}$ (Fig. 1D) decreases immediately away from half filling and develops a downward curvature toward $-1$. 
The value becomes slightly less negative with increasing doping, and the curvature reduces at around $x=-0.15$. 
In regimes where $\Gamma_d \bar{\chi_d}$ leans toward $-1$ and $\chi^{-1}_d$ simultaneously tends toward zero, the two diagnostics provide mutually reinforcing evidence for interaction-driven superconductivity. These trends persist at different $U$ values and system sizes (see Supplementary Figs.~\ref{fig:Pd_and_vertexU6}, \ref{fig:Pd_and_vertexU10}, \ref{fig:Pd_and_vertexU6_smaller}, \ref{fig:Pd_and_vertexU6_larger}).

\section*{Pattern for the pair-pair correlations}

To compare with prior studies, we directly visualize the real-space $d$-wave pair-pair correlations on $16\times 4$ lattices in Fig.~\ref{fig:realspacepattern}.
At half filling (Fig.~\ref{fig:realspacepattern} B), correlations are weak and short-ranged, consistent with gapped charge fluctuations. 
Upon hole doping (Fig.~\ref{fig:realspacepattern} A), the pattern departs from a simple $d_{x^2-y^2}$ sign structure and instead organizes into a ``plaquette''-like $d$-wave structure where the sign change of superconducting correlations happens around the cylinder instead of along the cylinder expecting for a 2D limit $d$-wave state, in close agreement with prior four-leg-cylinder DMRG results~\cite{doi:10.1126/science.aal5304,PhysRevB.102.041106}. This underscores that the $\tau=\beta/2$ probe targets low-energy physics.
By contrast, for electron doping (Fig.~\ref{fig:realspacepattern} C) we observe a conventional 2D $d_{x^2-y^2}$ pattern extending over several unit cells.
The pattern is disrupted at high temperature and strengthens upon cooling (see Supplementary Fig.~\ref{fig:realspace_pattern_Tdep}). 
The structure is robust over a finite doping window and evolves smoothly with further doping (see Supplementary Fig.~\ref{fig:real_space_doping_dep}), consistent with Fig.~\ref{fig:Pd_and_vertexU8} where $\chi^{-1}_d$ exhibits a progressively weaker decline upon cooling with increasing doping $|x|$.

Methodologically, choosing the unequal-time midpoint correlator $\tau= \beta/2$ suppresses high-energy contributions. Although finite temperature and the fermion sign limit the accessible correlation length, this choice reveals the spatial pattern much more clearly than the commonly used $\omega=0$ susceptibility, for which correlations decay far more rapidly and the contrast largely washes out (Supplementary Fig.~\ref{fig:real_space_doping_dep}).

Having benchmarked the rectangular geometry against prior DMRG, we next ask how the pattern behaves on a 2D square geometry, where DQMC is especially effective compared with DMRG. 
Figures~\ref{fig:realspacepattern} D, E show $8\times 8$ results, at a slightly higher temperature due to the stronger sign problem on the square geometry.
For hole doping (Fig.~\ref{fig:realspacepattern} D), the very local features resemble Fig.~\ref{fig:realspacepattern} A, but the plaquette $d$-wave arrangement loses coherence beyond short distances and no longer forms a clean ring-like pattern.
Electron doping (Fig.~\ref{fig:realspacepattern} E) retains the conventional 2D $d$-wave pattern over several unit cells, beyond which thermal effects weaken the $d$-wave pattern.
Overall, on a 2D geometry the conventional $d$-wave pattern persists for electron doping but not for hole doping.

\section*{Dependence on interaction strength}

We quantify the $U$ dependence via the dimensionless vertex $\Gamma_d \bar{\chi_d}$ in Fig.~\ref{fig:vertex_Udep}.
As $U$ increases, $\Gamma_d \bar{\chi_d}$ shifts downward toward $-1$ at fixed $T$.
The incremental change between $U/t=8$ and $10$ is modest, hinting at possible saturation over this range. These trends are weakly dependent on doping, 
and are qualitatively consistent with prior DCA results for  $t'=0$~\cite{PhysRevB.89.195133}. 
Complementary checks -- the contrasting temperature and doping dependence of $\chi_d$ between $U/t=0$ and $U/t=6$ (Supplementary Fig.~\ref{fig:non-U}), and the $U$-dependence of the real-space correlator (Supplementary Fig.~\ref{fig:realspace_Udep}) -- both indicate behavior qualitatively different from weakly interacting cases.
Together, these results strengthen the conclusion that the superconducting tendency in electron doping is interaction-driven and distinct from a weak-coupling scenario.

\section*{Quantifying the finite-temperature superconductivity tendency}

Despite the encouraging signatures of pairing, due to the fermion-sign problem, we remain well above $T_c$.
First, as $T\rightarrow T_c^+$, the long-time sector also controls the $\omega=0$ correlators, so the $\tau=\beta/2$ and $\omega=0$ correlators are expected to have consistent behavior (see Supplementary Formalism Section). 
In sign-problem-free DQMC for the attractive-$U$ Hubbard model, we verified that the $\tau=\beta/2$, $\omega=0$, and equal-time~\cite{PhysRevB.69.184501} susceptibilities exhibit consistent temperature scaling and yield the same $T_c$.
In our accessible temperature range these measures remain distinct (also see Ref.~\cite{jpsj} for temperature dependence of $\chi_d(\omega=0)$), indicating that we have not yet entered the asymptotic critical window.
Second, close to the transition temperature we expect $\chi_d$ to grow strongly with system size as pairing correlations expand toward the boundaries. In our data, over the temperatures and sizes we can reach, $\chi_d$ shows only weak size dependence throughout the bulk of the doping window with $d$-wave correlations (see Supplementary Fig.~\ref{fig:size_dep_elec}), again implying $T>T_c$ and that extracting $T_c$ from the present data would be unreliable.

One can nevertheless quantify the finite-temperature tendency toward superconductivity without assuming a specific critical form. We visualize the negative temperature slope of the pair susceptibility, $-\partial \chi_d/\partial T$, as a colormap in Fig.~\ref{fig:slope};
this captures how rapidly pairing fluctuations build up upon cooling.
The signal is small on the hole-doped side but sizable over a substantial portion of the electron-doped regime down to our lowest $T$. The map exhibits a dome-like region with a maximum near $1/8$ electron doping. Around this doping, we also see the earliest signs of increasing size sensitivity (see Supplementary Fig.~\ref{fig:size_dep_elec}), suggesting a possibly enhanced characteristic scale, although a controlled determination of $T_c$ remains out of reach.

\section*{Discussion and outlook}

Our results reveal a sharp contrast between hole- and electron-doped regimes: clear interaction-driven $d$-wave pairing tendencies appear on the electron-doped side, whereas the hole-doped side shows no clear indication of superconductivity upon cooling over the accessible temperatures.
This trend differs from cuprate experiments and from some of the recent numerical studies~\cite{doi:10.1126/science.adh7691}, underscoring the importance of the present analysis in clarifying our understanding of superconductivity in the Hubbard model.

A key methodological point is our emphasis on the $\tau=\beta/2$ method. This method has been underutilized in finite-temperature numerical studies of superconducting and other instabilities.
By construction, evaluating correlators at $\tau=\beta/2$ filters high-frequency components and emphasizes low-energy contributions in a controlled, numerically exact way, without notoriously ill-posed analytic continuation~\cite{jarrell1996bayesian}, and without additional assumptions that can bias outcomes.
This control is particularly valuable in the Hubbard model, where strongly competing orders make robustness and exactness essential.

The discrepancy between our model trends and experiments likely reflects physics beyond the minimal Hubbard model. 
Candidate ingredients include multi-orbital physics, electron-phonon coupling, disorder, and longer-range hoppings and interactions.
Crucially, the $\tau=\beta/2$ method gives a practical path to probe such extensions even as increasing model complexity may worsen the sign problem: it keeps low-energy access at finite 
$T$ while retaining numerical exactness.
In this sense, algorithms such as DQMC can be more powerful than often assumed, despite the fermion sign problem.

Looking forward, we expect our work to motivate refinements of theoretical tools that improve the extraction of $T_c$. Moreover, using these finite-$T$ benchmarks to cross-calibrate methods that reach lower temperatures or the ground state will help build a coherent picture across temperature scales, establishing a stable foundation for assessing which extensions to the Hubbard model are needed to understand the high-$T_c$ cuprates.

\bibliography{main}
\bibliographystyle{Science}

\section*{Acknowledgements}
We acknowledge helpful discussions with L.~Balents, E.~W.~Huang, H.-C.~Jiang, S.~Jiang, S.~A.~Kivelson, R.~B.~Laughlin, B.~Moritz, R.~T.~Scalettar, and J.-X.~Zhang.
W.O.W. acknowledges support from the Gordon and Betty Moore Foundation through Grant GBMF8690 to the University of California, Santa Barbara, to the Kavli Institute for Theoretical Physics (KITP).
T.P.D. acknowledges support from the U.S. Department of Energy (DOE), Office of Basic Energy Sciences,
Division of Materials Sciences and Engineering. 
Computational work was performed on the Sherlock cluster at Stanford University and on resources of the National Energy Research Scientific Computing Center, a US Department of Energy, Office of Science User Facility, using NERSC award BES-ERCAP0031425.
This research was supported in part by grant NSF PHY-2309135 to the KITP.
The data and analysis routines needed to reproduce the figures, as well as the DQMC source code, can be found at \url{https://doi.org/10.5281/zenodo.17353680}.

\section*{Supplementary Materials}
Supplementary Text\\
Figs.~S1 to S10\\
Reference (\textit{40})

\clearpage

\begin{figure}[tbp]
    \centering
    \includegraphics[width=\linewidth]{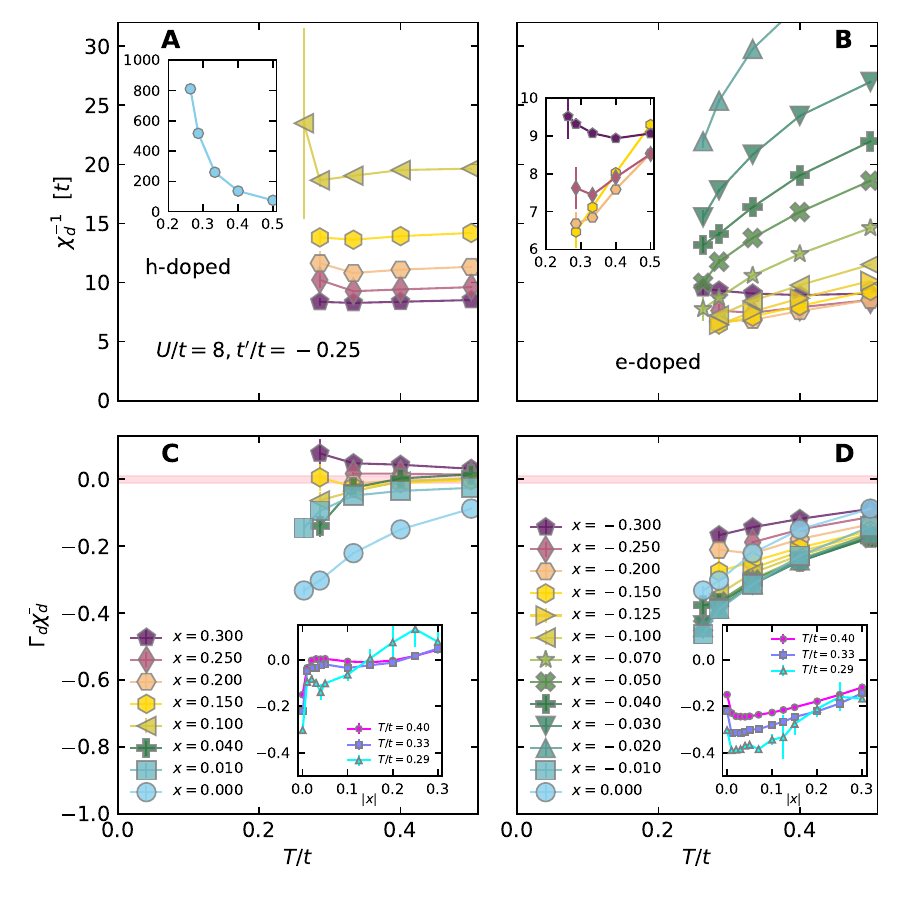}
    \caption{ (A,B) Unequal-time $d$-wave pair-field susceptibility, $\chi_d \equiv \beta\langle \mathcal{T}_\tau\Delta_d(\tau=\beta/2)\Delta^\dagger_d\rangle/N_s$ for hole doping ($x > 0 $; A) and electron doping ($x < 0$; B). The inset of (A) highlights half filling ($x=0$); the inset of (B) zooms in on the electron-doped regime $|x| \geq 15\%$. (C, D) Superconducting pairing vertex from the unequal-time susceptibility and its uncorrelated bubble, $\Gamma_d \bar{\chi_d} \equiv \bar{\chi_d}/ \chi_d -1$ (with $\bar{\chi_d}$ built from single-particle Green’s functions), for hole doping (C) and electron doping (D). Insets in (C, D) show the doping dependence of $\Gamma_d \bar{\chi_d}$ at fixed temperatures.
    Parameters: $U/t=8$, $t'/t=-0.25$; cluster $8\times 8$. 
}
    \label{fig:Pd_and_vertexU8}
\end{figure}

\begin{figure}[tbp]
    \centering
    \includegraphics[width=\linewidth]{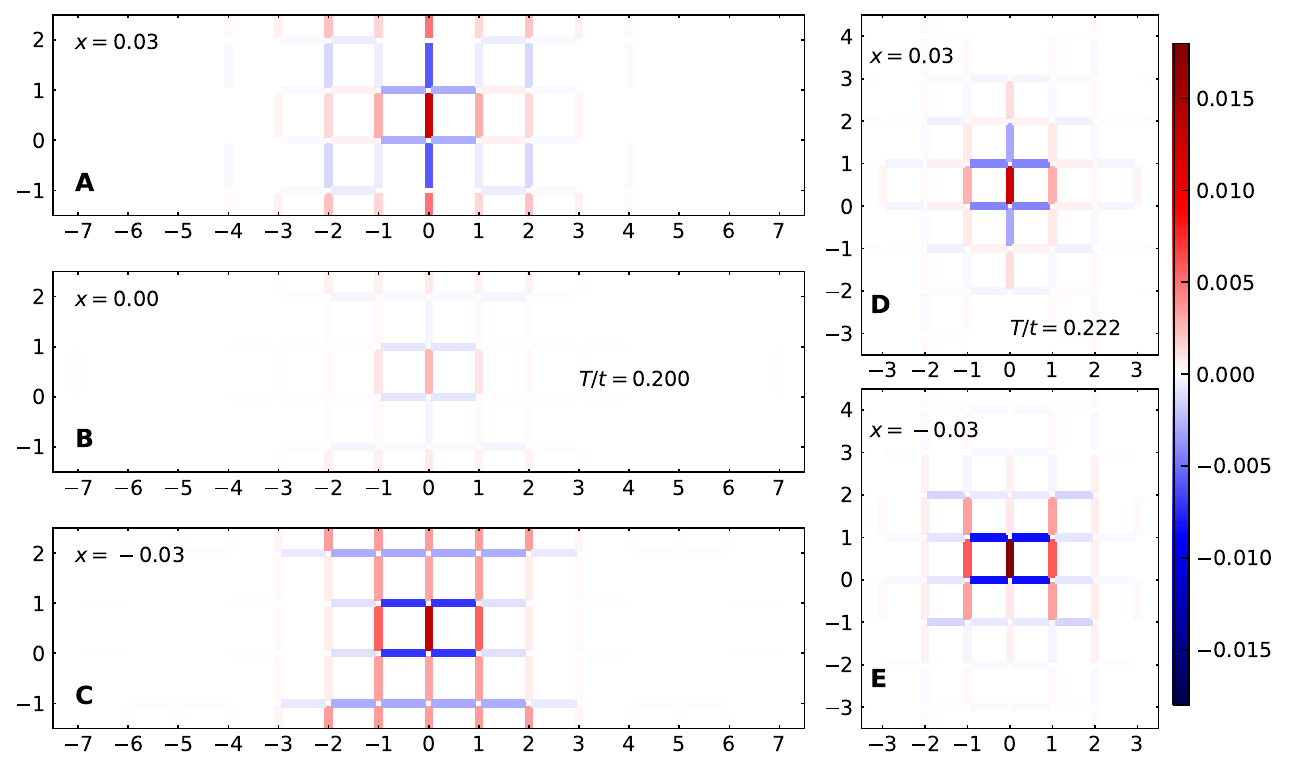}
    \caption{ 
    Real-space pattern of the unequal-time pair-pair correlator
    $\beta S_{\alpha,\alpha'}(\mathbf{r},\tau=\beta/2) \equiv \beta \sum_i \big\langle  \mathcal{T}_\tau \Delta_{\alpha}(\tau=\beta/2, \mathbf{r}_i + \mathbf{r}) \Delta_{\alpha'}^\dagger(0, {\mathbf{r}_i}) \big\rangle/{N_s} $ (see Supplementary Materials Formalism Section), at (A, D) hole doping 
$x=0.03$, (B) half filling $x=0$, and (C, E) electron doping 
$x=-0.03$. 
We take the vertical bond connecting $(0,0)$ and $(0,1)$ as the reference.
Parameters: $U/t=6$, $t'/t=-0.25$; 
Temperatures and sizes: $T/t=0.2$ on $16\times 4$ (A-C); $T/t=0.222$ on $8\times 8$ (D, E).
}
    \label{fig:realspacepattern}
\end{figure}

\begin{figure}[tbp]
    \centering
    \includegraphics[width=\linewidth]{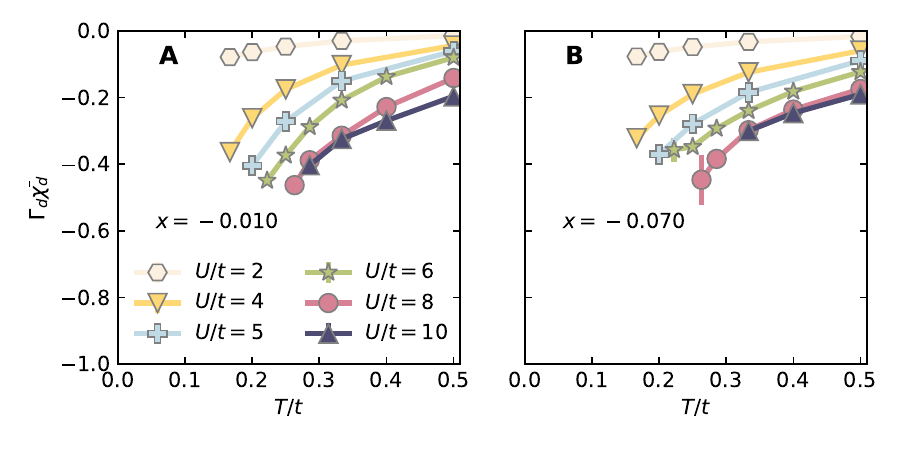}
    \caption{ Superconducting pairing vertex $\Gamma_d \bar{\chi_d}$ for various interaction strengths $U$ at electron doping $x=-1\%$ (A) and $x=-7\%$ (B). Parameters: $t'/t=-0.25$; cluster $8\times 8$. 
}
    \label{fig:vertex_Udep}
\end{figure}

\begin{figure}[tbp]
    \centering
    \includegraphics[width=\linewidth]{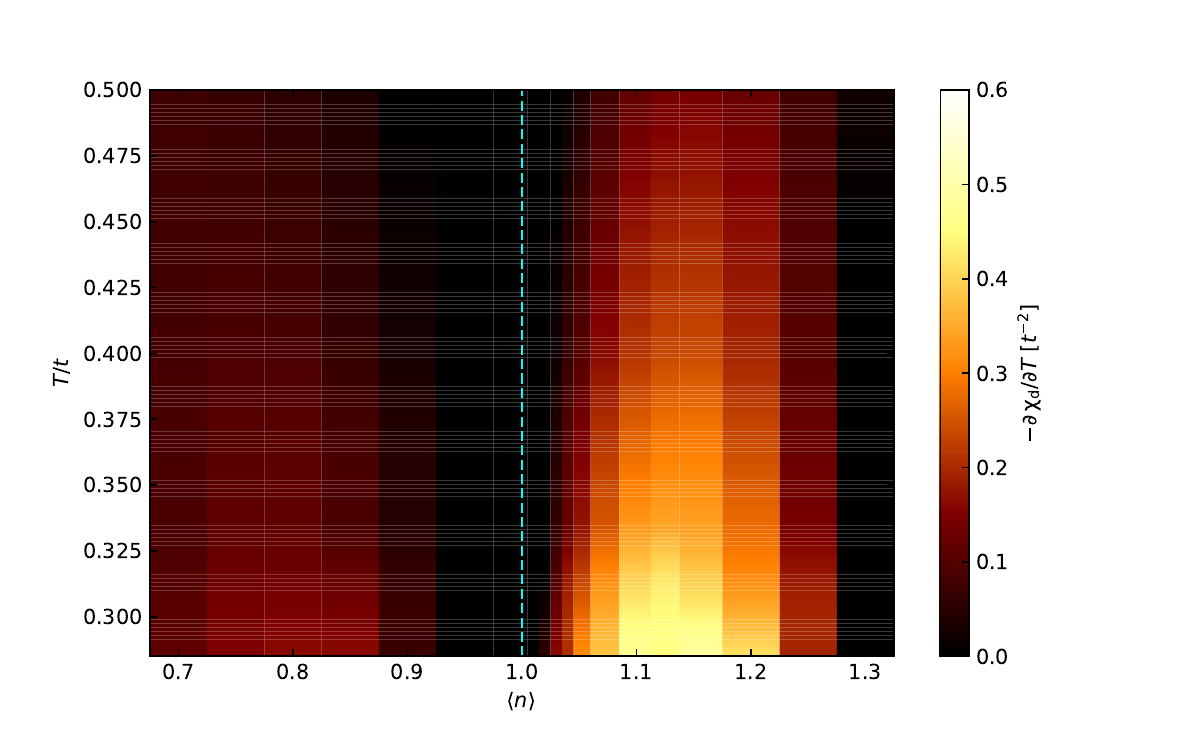}
    \caption{ Negative temperature slope of the pair susceptibility, $-\partial \chi_d/\partial T$, estimated from an Akima spline fit to $\chi_d(T)$.
    Parameters: $U/t=6$, $t'/t=-0.25$; cluster $8\times 8$.
}
    \label{fig:slope}
\end{figure}

\clearpage
\renewcommand{\thefigure}{S\arabic{figure}}
\setcounter{figure}{0}

\section*{Supplementary Materials}

\subsection*{Supplementary Text}

\noindent\underline{Error Analysis} \label{sec:err}

Error bars denote statistical uncertainties and were calculated using jackknife resampling. In real-space maps of the correlation functions $S_{\alpha,\alpha'}(\mathbf{r})$, bonds are colored only when $|S_{\alpha,\alpha'}(\mathbf{r})| > 2\sigma$, with $\sigma$ the jackknife error.

\noindent\underline{Formalism} \label{sec:formalism}

For convenience, $k_B$ and $\hbar$ are set to $1$ throughout the paper.
We simulate the $t$-$t'$-$U$ Hubbard model on square or rectangular lattices with periodic boundary conditions in both spatial directions. The model is translation invariant.
In imaginary time $\tau$ and real space $\mathbf{r}$, the singlet-pair correlator is
\begin{equation}
    S_{\alpha,\alpha'}(\mathbf{r},\tau) = \frac{1}{N_s} \sum_i \big\langle  \mathcal{T}_\tau \Delta_{\alpha}(\tau, \mathbf{r}_i + \mathbf{r}) \Delta_{\alpha'}^\dagger(0, {\mathbf{r}_i}) \big\rangle , \label{corr_def}
\end{equation}
with bond-pair creation operator
\begin{equation}
    \Delta_{\alpha}^\dagger(\mathbf{r}_i) = \frac{1}{\sqrt{2}} \! \left(c_{\mathbf{r}_i,\uparrow}^\dagger c_{\mathbf{r}_i+\alpha,\downarrow}^\dagger - c_{\mathbf{r}_i,\downarrow}^\dagger c_{\mathbf{r}_i+\alpha,\uparrow}^\dagger\right), \qquad \alpha \in \{\hat{\mathbf x},\hat{\mathbf y}\},
\end{equation}
$\mathcal{T}_\tau $ the imaginary-time ordering operator, and $N_s$ the number of lattice sites.
Writing $K \equiv H-\mu N$ (with $H$ the $t$-$t'$-$U$ Hubbard Hamiltonian, $\mu$ the chemical potential, and $N$ the particle-number operator) and $\Delta_\alpha(\tau, \mathbf{r})=e^{\tau K} \Delta_\alpha(\mathbf{r})e^{-\tau K}$, we have
\begin{align}
    S_{\alpha,\alpha'}(\mathbf{r},\tau) =  \frac{1}{N_s Z} \sum_i\Tr\! \left[e^{-\beta K} \, \mathcal{T}_\tau  \Delta_{\alpha}(\tau, \mathbf{r}_i + \mathbf{r}) \Delta_{\alpha'}^\dagger(0, {\mathbf{r}_i}) \right], \label{traceeq}
\end{align}
where $Z=\Tr[e^{-\beta K}]$ is the grand-canonical partition function, and $\Tr$ denotes taking the trace.
For $0<\tau<\beta$, the Lehmann representation is
\begin{align}
  S_{\alpha,\alpha'}(\mathbf{r},\tau)= \frac{1}{N_s Z}\sum_{m,n,i}
    \langle n|\Delta_{\alpha} (\mathbf r_i+\mathbf r) |m\rangle
    \langle m| \Delta_{\alpha'}^\dagger(\mathbf r_i)|n\rangle\,
    e^{-\beta E_n}\,e^{\tau(E_n-E_m)},                        \label{eq:corr_tau_lemm}
\end{align}
with $K|n\rangle = E_n |n\rangle$.
By cyclicity of the trace, the correlator is periodic in imaginary time
\begin{align}
S_{\alpha,\alpha'}(\mathbf{r},\tau-\beta) = S_{\alpha,\alpha'}(\mathbf{r},\tau).
\end{align}
The equal-time contact discontinuity at the endpoints is
\begin{align}
S_{\alpha,\alpha'}(\mathbf{r},\tau=0^+)-S_{\alpha,\alpha'}(\mathbf{r},\tau=\beta^-)=\frac{1}{N_s}\sum_i \Big\langle \big[\Delta_{\alpha}(\mathbf{r}_i + \mathbf{r}), \Delta_{\alpha'}^\dagger({\mathbf{r}_i}) \big]\Big\rangle,
\end{align}
where $[..,..]$ denotes a commutator.

The $d_{x^2-y^2}$-wave pair-creation operator is 
\begin{align}
\Delta^\dagger_d & = \frac{1}{\sqrt{2}} \! \sum_{\mathbf{k}} \left(\cos k_x - \cos k_y \right) \frac{1}{\sqrt{2}}\left(c^\dagger_{\mathbf{k},\uparrow} c^\dagger_{-\mathbf{k},\downarrow} - c^\dagger_{\mathbf{k},\downarrow} c^\dagger_{-\mathbf{k},\uparrow}\right)\nonumber\\
& =  \frac{1}{2} \! \sum_i \left[ \left(c_{\mathbf{r}_i,\uparrow}^\dagger c_{\mathbf{r}_i+\hat{\mathbf{x}},\downarrow}^\dagger - c_{\mathbf{r}_i,\downarrow}^\dagger c_{\mathbf{r}_i+\hat{\mathbf{x}},\uparrow}^\dagger\right) - \left(c_{\mathbf{r}_i,\uparrow}^\dagger c_{\mathbf{r}_i+\hat{\mathbf{y}},\downarrow}^\dagger - c_{\mathbf{r}_i,\downarrow}^\dagger c_{\mathbf{r}_i+\hat{\mathbf{y}},\uparrow}^\dagger\right)
\right] \nonumber \\
& = \frac{1}{\sqrt{2}} \sum_i \left( \Delta_{\hat{\mathbf{x}}}^\dagger(\mathbf{r}_i) -  \Delta_{\hat{\mathbf{y}}}^\dagger(\mathbf{r}_i)  \right), \label{eq:deltaD}
\end{align}
where we used the Fourier transform
\begin{align}
c^\dagger_{\mathbf{k},\sigma} = \frac{1}{\sqrt{N_s}} \sum_{i} e^{i\mathbf{k}\cdot \mathbf{r}_i} c^\dagger_{\mathbf{r}_i,\sigma}.
\end{align}
The unequal-time susceptibility is given as
\begin{align}
\chi_d(\tau) = \frac{\beta}{N_s}\langle \mathcal{T}_\tau \Delta_d (\tau) \Delta^\dagger_d \rangle, \label{eq:defchid}
\end{align}
and we focus on $\tau=\beta/2$.
Using Eq.~\eqref{corr_def}, this can be written as
\begin{align}
\chi_d(\tau) = \frac{\beta}{2} \sum_i \left[ S_{\hat{\mathbf{x}},\hat{\mathbf{x}}}(\mathbf{r}_i,\tau) + S_{\hat{\mathbf{y}},\hat{\mathbf{y}}}(\mathbf{r}_i,\tau) - 
S_{\hat{\mathbf{x}},\hat{\mathbf{y}}}(\mathbf{r}_i,\tau) -
S_{\hat{\mathbf{y}},\hat{\mathbf{x}}}(\mathbf{r}_i,\tau) \right]. \label{eq:combineStoChi}
\end{align}
From Eq.~\eqref{eq:defchid}, the Lehmann representation for $0<\tau<\beta$ is
\begin{align}
  \chi_d(\tau)= \frac{\beta}{N_s Z}\sum_{m,n}
    \langle n|\Delta_d |m\rangle
    \langle m| \Delta_d^\dagger|n\rangle\,
    e^{-\beta E_n}\,e^{\tau(E_n-E_m)}.                        \label{eq:chid_tau_lemm}
\end{align}
Throughout the paper we use the shorthand $\chi_d\equiv\chi_d(\tau=\beta/2)$. This differs from the more common $\chi_d(\omega=0)$, which is often closer to experimental probes, and from $\chi_d(\tau=0^+)$, which is convenient in methods without dynamics (e.g., DMRG).
We choose $\tau=\beta/2$ because it emphasizes low-energy physics. 
More specifically but still at an intuitive level, from the spectral/Lehmann forms, Eqs.~\eqref{eq:corr_tau_lemm} and \eqref{eq:chid_tau_lemm}, for $S_{\alpha,\alpha'}(\mathbf{r},\tau)$ and $\chi_d(\tau)$, contributions with energy difference $\Delta E$ enter with factors $e^{-\tau\Delta E}$  and $e^{-(\beta-\tau)\Delta E}$. Taking $\tau=\beta/2$ treats the forward and backward time sectors symmetrically and suppresses high-energy contributions, causing the low-energy structure to stand out.

In what follows, we present a more detailed and mathematically precise justification for why $\tau=\beta/2$ is better for probing low-energy physics than other choices. The analysis is presented for $\chi_d(\tau)$, but it generalizes directly to the real-space, bond-resolved correlators $S_{\alpha,\alpha'}(\mathbf{r},\tau)$. Related material appears in the Supplementary Materials of Ref.~\cite{chen2025topological}.

We adopt the following convention for the susceptibility in Matsubara frequency space (bosonic), $\omega_\nu=2\pi\nu/\beta$:
\begin{equation}
  \chi_d(i\omega_\nu)=\frac{1}{\beta}\!\int_0^\beta\!\!d\tau\,\chi_d(\tau)e^{i\omega_\nu\tau}. \label{eq:fourieromega}
\end{equation}
This convention offsets the prefactor $\beta$ introduced in Eq.~\eqref{eq:defchid}.
Using Eq.~\eqref{eq:chid_tau_lemm}, 
\begin{equation}
  \chi_d(i\omega_\nu) =-\frac{1}{N_s Z}\sum_{m,n}\!
      \frac{|\langle m|\Delta^\dagger_d|n\rangle|^{2}\bigl(e^{-\beta E_n}-e^{-\beta E_m}\bigr)}
           {i\omega_\nu+E_n-E_m}.                                   \label{eq:chi_iwn}
  \end{equation}
Analytic continuation $i\omega_\nu\to\omega+i0^+$ gives
\begin{equation}
  \chi_d(\omega)=
  -\frac{1}{N_s Z}\sum_{m,n}
      \frac{|\langle m|\Delta^\dagger_d|n\rangle|^{2}\bigl(e^{-\beta E_n}-e^{-\beta E_m}\bigr)}
           {\omega+E_n-E_m+i0^+}.                                  \label{eq:chi_omega}
\end{equation}

Using $1/(x+i0^+)=\mathcal P(1/x)-i\pi\delta(x)$, the imaginary part is
\begin{equation}
  \Im\chi_d(\omega)=\frac{\pi}{ N_s Z}\sum_{m,n}
     |\langle m|\Delta^\dagger_d|n\rangle|^{2}\bigl(e^{-\beta E_n}-e^{-\beta E_m}\bigr)\,
     \delta\!\left(\omega-(E_m-E_n)\right).                       \label{eq:Im_chi}
\end{equation}
Hence $\Im\chi_d(\omega)$ directly resolves the excitation energies $\omega=E_m-E_n$ that couple to $\Delta^\dagger_d$. In other words, the low-energy information is embedded in the low-energy spectrum in $\Im\chi_d(\omega)$.

Combining Eqs.~\eqref{eq:chid_tau_lemm} and \eqref{eq:Im_chi} yields
\begin{equation}
  \chi_d(\tau)=\int_{-\infty}^{\infty}\!\beta d\omega\;
     W_\tau(\omega)\,
     \frac{\Im\chi_d(\omega)}{\pi},                          \label{eq:chi_tau_kernel}
\end{equation}
with the effective kernel,
\begin{equation}
  W_\tau(\omega)=\frac{e^{-\tau\omega}}{1-e^{-\beta\omega}}
 =\frac{e^{-(\tau-\frac{\beta}{2})\omega}}{2\sinh(\beta\omega/2)}.             \label{eq:kernel}
\end{equation}
Setting $\tau=\beta/2$ gives
\begin{equation}
  \chi_d\!\left(\tau=\tfrac{\beta}{2}\right)
  =\int_{-\infty}^{\infty}\!\beta d\omega\;
     \frac{1}{2\sinh(\beta\omega/2)}\,
     \frac{\Im\chi_d(\omega)}{\pi}.                                 \label{eq:chi_beta_half}
\end{equation}
The kernel $[2\sinh(\beta\omega/2)]^{-1}$ places strong weight near $\omega=0$ and decays exponentially for $|\beta\omega|\gtrsim2$.
For general $\tau$ and $|\beta\omega|\ll 1$, the low-frequency expansion of $W_\tau(\omega)$ reads
\[
  W_\tau(\omega)=\frac{1}{\beta\omega} + \frac{\left(\frac{\beta}{2}-\tau\right)}{\beta} + \frac{\left(\frac{\beta}{2}-\tau\right)^2\omega}{2\beta}-\frac{\beta \omega}{24} + \mathcal O(\omega^{2}),
\]
so $\tau=\beta/2$ removes the $\mathcal{O}(\omega^0)$ term.

From Eqs.~\eqref{eq:chi_iwn} and \eqref{eq:Im_chi}, the $\omega=0$ susceptibility is
\begin{equation}
  \chi_d(\omega=0)=\int_{-\infty}^{\infty}\!\beta d\omega\;
     \frac{1}{\beta\omega}\,
     \frac{\Im\chi_d(\omega)}{\pi},  
     \label{eq:chi_omega_zero}
\end{equation}
whose algebraic kernel $(\beta\omega)^{-1}$ suppresses high-frequency contributions only as a power law.  Equation~\eqref{eq:chi_beta_half} shares the same low-frequency $(\beta\omega)^{-1}$ behavior, but damps the high-frequency features exponentially and is a more selective probe of low-energy physics.

Alternatively, we can analyze the Lehmann forms directly. From Eqs.~\eqref{eq:chid_tau_lemm} and \eqref{eq:chi_omega}, both $\chi_d(\tau=\beta/2)$ and $\chi_d(\omega=0)$ can be written as
\begin{align}
\frac{\beta}{N_s Z}\sum_{m,n} \big|\langle m|\Delta^\dagger_d|n\rangle\big|^{2} \mathcal{F}_X(\beta,E_n,E_m),
\end{align}
with
\begin{align}
\mathcal{F}_{X=(\tau=\beta/2)}(\beta,E_n,E_m) &= e^{-\frac{\beta E_n}{2}-\frac{\beta E_m}{2}}=e^{-\beta E_m} e^{-\frac{\beta(E_n-E_m)}{2}}, \\
\mathcal{F}_{X=(\omega=0)}(\beta,E_n,E_m) &= \frac{e^{-\beta E_n}-e^{-\beta E_m}}{\beta(E_m-E_n)}=e^{-\beta E_m} \frac{e^{-\beta (E_n-E_m)}-1}{\beta(E_m-E_n)}.
\end{align}
Note that for a number-conserving Hamiltonian, matrix elements with $m=n$ vanish for a pair operator, so the apparent $0/0$ in $\chi_d(\omega=0)$ at $E_m=E_n$  is not an issue.

When the dominant contributions come from near-degenerate pairs $(m,n)$ with
\begin{align}
x\equiv \beta(E_n-E_m) \text{ satisfying } |x|\ll1
\end{align}
(e.g., in the critical regime near a finite $T_c$, where the characteristic pairing-fluctuation scale $\Omega_{\text{pair}}(T)$ obeys $\beta \Omega_{\text{pair}}\ll 1$), it is justified to expand $\mathcal{F}_X$ in powers of $x$:
\begin{align}
\mathcal{F}_{X=(\tau=\beta/2)}&=&e^{-\beta E_m} \left[ 1-\frac{1}{2}x+ \frac{1}{8}x^2+ \mathcal{O}(x^3)\right], \label{betaover2F}\\
\mathcal{F}_{X=(\omega=0)} &=&e^{-\beta E_m} \left[1-\frac{1}{2}x + \frac{1}{6} x^2+ \mathcal{O}(x^3)\right]. \label{omega0F}
\end{align}
The leading two terms coincide, implying that as low-energy physics dominates and $|x|\rightarrow 0$, the two susceptibilities converge. 
The first difference appears at order $x^2$: the coefficient is smaller for $\tau=\beta/2$ ($1/8<1/6$), indicating that, at fixed $|x|$, higher-energy corrections enter $\chi_d(\omega=0)$ more strongly than $\chi_d(\tau=\beta/2)$.
Equivalently, the $\tau=\beta/2$ definition assigns smaller weight to transitions with finite $x$, and therefore suppresses high-energy contributions more effectively.
This again shows that, in practice (and especially under fermion-sign constraints), $\chi_d(\tau=\beta/2)$ is more low-energy-biased, while at sufficiently low temperatures the two definitions are expected to agree.

\noindent\underline{Supplementary data for superconductivity susceptibility, vertex, and real-space pairing correlations}

\begin{figure}[tbp]
    \centering
    \includegraphics[width=\linewidth]{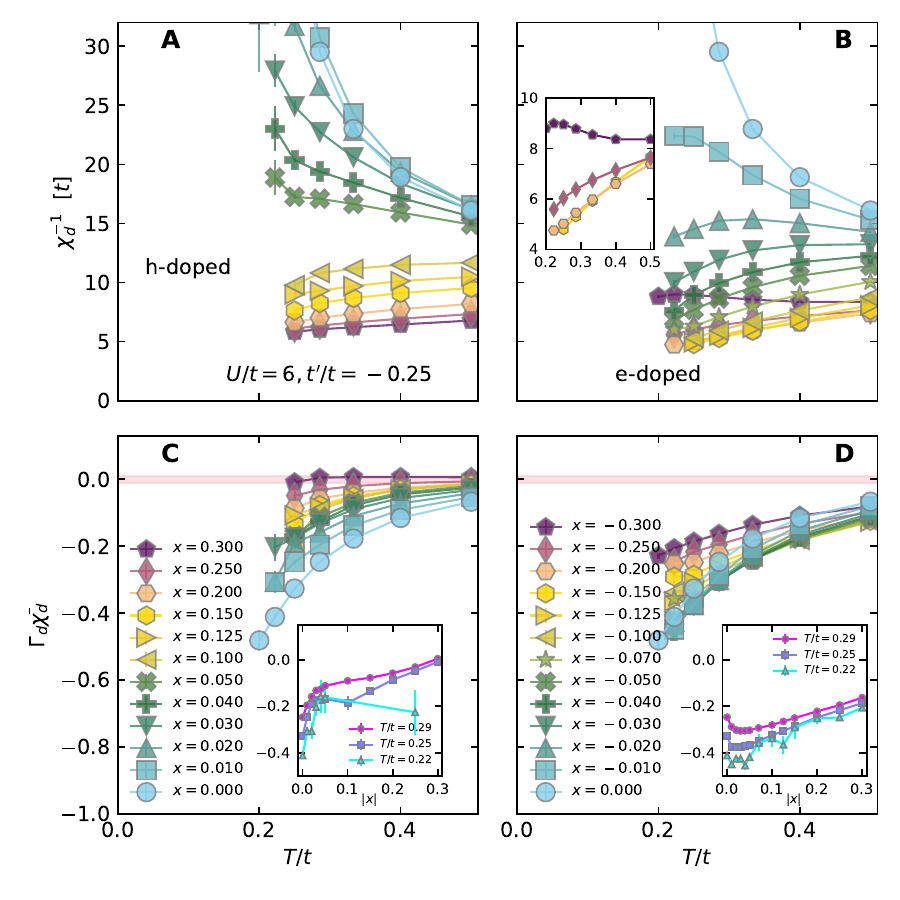}
    \caption{ Unequal-time $d$-wave pair-field susceptibility and vertex (definitions as in main-text Fig.~\ref{fig:Pd_and_vertexU8}), now for $U/t=6$ (other parameters unchanged). 
    Trends are qualitatively consistent with the $U/t=8$ case in main-text Fig.~\ref{fig:Pd_and_vertexU8}.
}
    \label{fig:Pd_and_vertexU6}
\end{figure}

\begin{figure}[tbp]
    \centering
    \includegraphics[width=\linewidth]{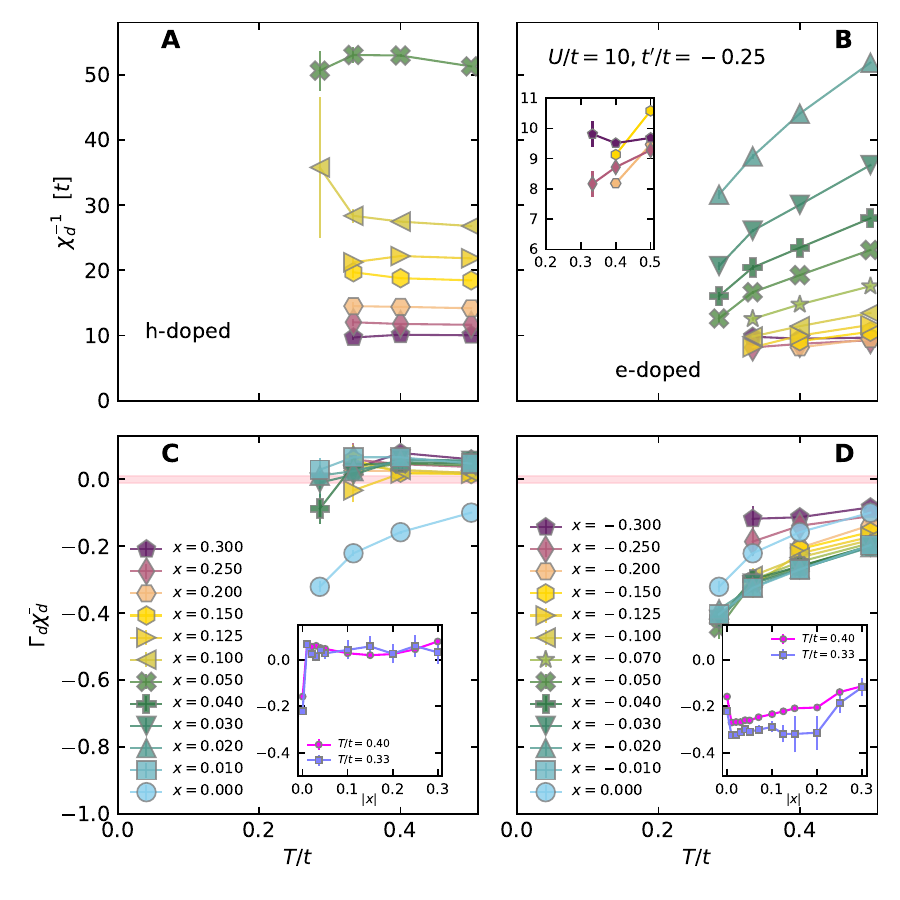}
    \caption{ Unequal-time $d$-wave pair-field susceptibility and vertex (definitions as in main-text Fig.~\ref{fig:Pd_and_vertexU8}), now for $U/t=10$ (other parameters unchanged).
    Trends are qualitatively consistent with the $U/t=8$ case in main-text Fig.~\ref{fig:Pd_and_vertexU8}.
}
    \label{fig:Pd_and_vertexU10}
\end{figure}

Figures~\ref{fig:Pd_and_vertexU6} and \ref{fig:Pd_and_vertexU10} present the same analysis as Fig.~\ref{fig:Pd_and_vertexU8} for $U/t=6$ and $U/t=10$; the results are qualitatively consistent.

\begin{figure}[tbp]
    \centering
    \includegraphics[width=\linewidth]{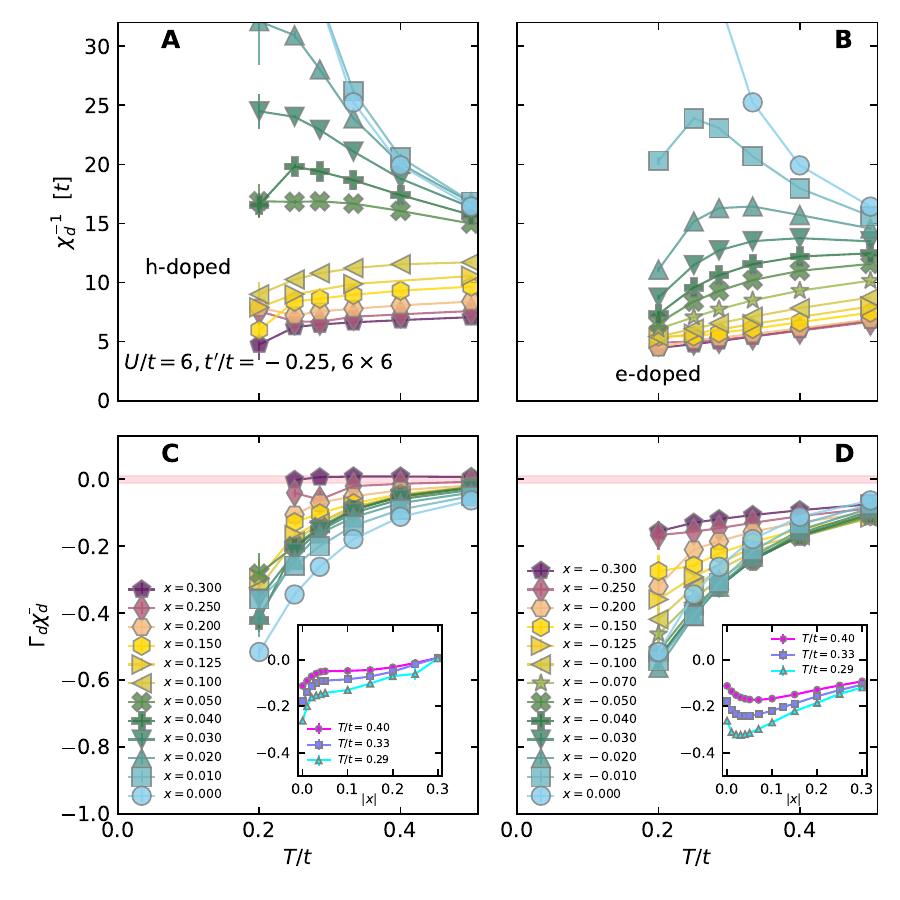}
    \caption{ Same analysis as Fig.~\ref{fig:Pd_and_vertexU6}: $d$-wave pair-field susceptibility and vertex for $U/t=6$, but on a $6\times 6$ lattice (other parameters as in Fig.~\ref{fig:Pd_and_vertexU6}). 
    Trends are qualitatively consistent with the $8 \times 8$ case in Fig.~\ref{fig:Pd_and_vertexU6}.
}
    \label{fig:Pd_and_vertexU6_smaller}
\end{figure}

\begin{figure}[tbp]
    \centering
    \includegraphics[width=\linewidth]{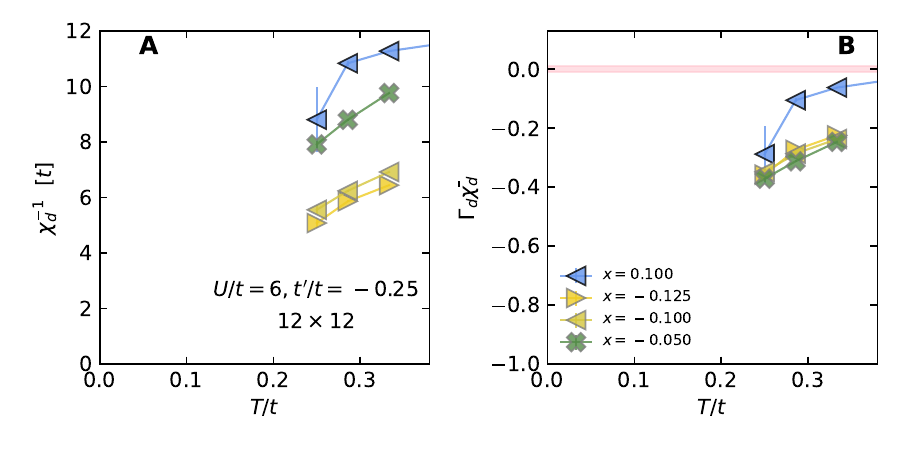}
    \caption{ (A) $d$-wave pair-field susceptibility and (B) superconducting pairing vertex for $U/t=6$ on a $12\times 12$ lattice (other parameters as in Fig.~\ref{fig:Pd_and_vertexU6}). 
    Trends are qualitatively consistent with the $8 \times 8$ case in Fig.~\ref{fig:Pd_and_vertexU6}.
}
    \label{fig:Pd_and_vertexU6_larger}
\end{figure}

Figures~\ref{fig:Pd_and_vertexU6_smaller} and \ref{fig:Pd_and_vertexU6_larger} show the same analysis as Fig.~\ref{fig:Pd_and_vertexU6}, now for sizes $6\time 6$ and 
$12\times 12$ (other parameters unchanged). The trends are qualitatively consistent with the $8\times 8$case in Fig.~\ref{fig:Pd_and_vertexU6}.

\begin{figure}[tbp]
    \centering
    \includegraphics[width=\linewidth]{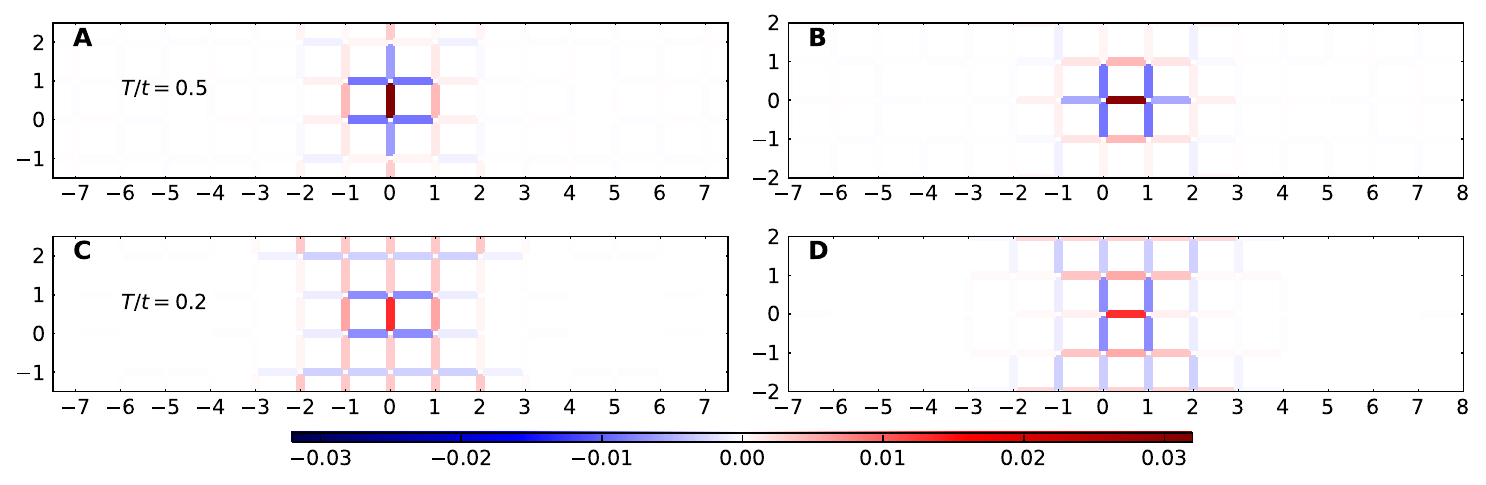}
    \caption{ Pair-pair correlation function $\beta S_{\alpha,\alpha'}(\mathbf{r},\tau=\beta/2)$ at $x=-0.03$ for two temperatures and two reference-bond orientations ($U/t=6$, $t'/t=-0.25$; lattice size $16\times 4$).
    Top: higher $T/t=0.5$; bottom: lower $T/t=0.2$. Left: horizontal reference bond; right: vertical reference bond.
}
    \label{fig:realspace_pattern_Tdep}
\end{figure}

Figure~\ref{fig:realspace_pattern_Tdep} shows $\beta S_{\alpha,\alpha'}(\mathbf{r},\tau=\beta/2)$ at fixed electron doping $x=-0.03$ for two temperatures and for horizontal vs vertical choices of the reference bond.
High temperature (A, B) shows a non-$d$-wave pattern with short-range correlations, whereas lower temperature (C, D) reveals a $d$-wave pattern with more extended real-space correlations.
Choosing a vertical reference bond (B, D) yields the same qualitative behavior as choosing a horizontal one (A, C).

\begin{figure}[tbp]
    \centering
    \includegraphics[width=0.9\linewidth]{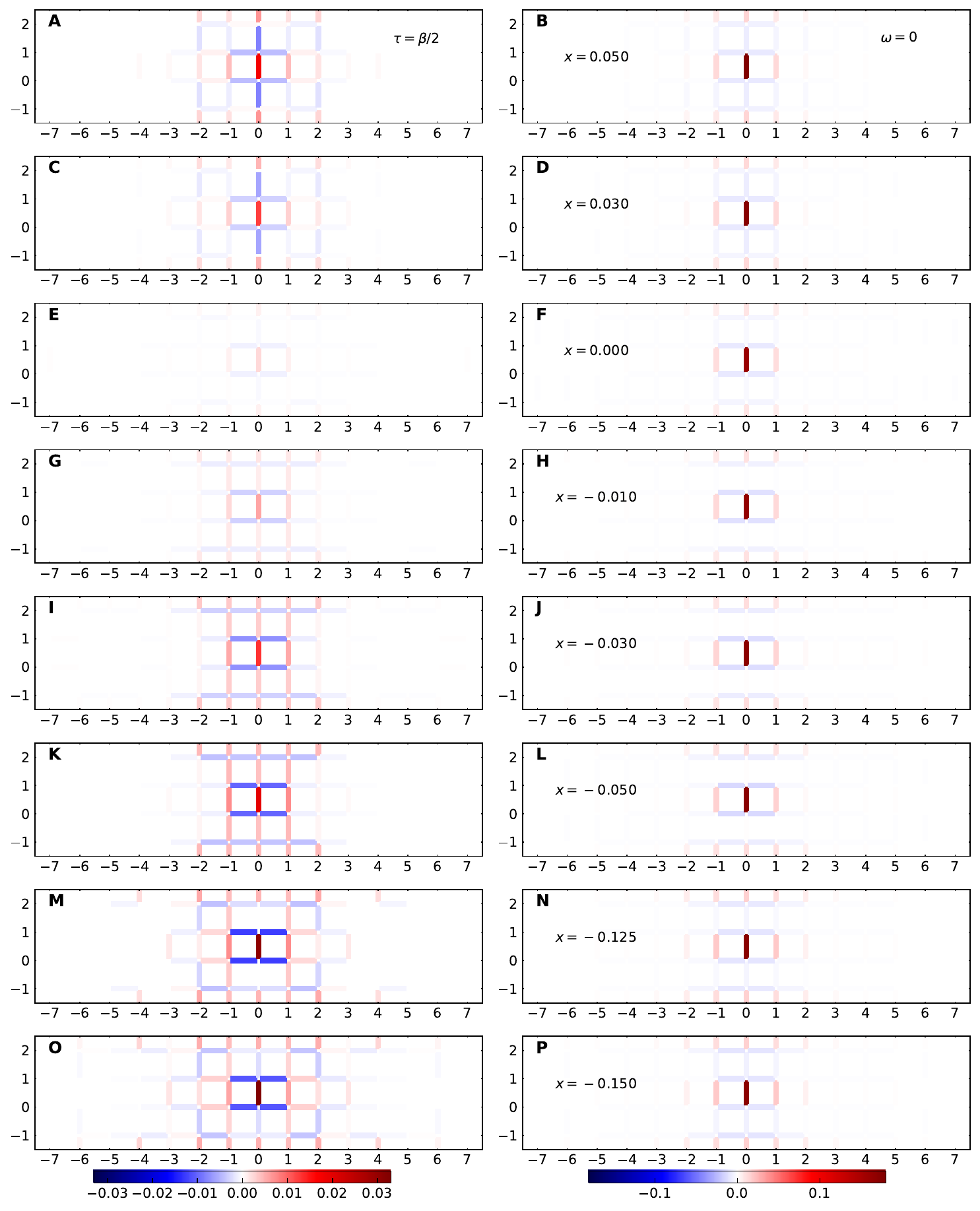}
    \caption{ Pair-pair correlation function $\beta S_{\alpha,\alpha'}(\mathbf{r},\tau=\beta/2)$ (left) and $ S_{\alpha,\alpha'}(\mathbf{r},\omega=0)\equiv \int_0^\beta d\tau S_{\alpha,\alpha'}(\mathbf{r},\tau) $ (right) at $T/t=0.2$ for varying $x$ (same $x$ per row; $U/t=6$, $t'/t=-0.25$; lattice size $16\times 4$).
    Here the $\tau \rightarrow \omega$ transform follows the standard Matsubara convention and does not include the additional $1/\beta$ prefactor used in Eq.~\eqref{eq:fourieromega}. 
    We take the same vertical bond as in main-text Fig~\ref{fig:realspacepattern} as the reference.
}
\label{fig:real_space_doping_dep}
\end{figure}

Figure~\ref{fig:real_space_doping_dep} (on the left) shows the doping evolution of $\beta S_{\alpha,\alpha'}(\mathbf{r},\tau=\beta/2)$ at fixed temperature. At half filling the correlations are very weak. Hole doping yields  a plaquette $d$-wave structure, consistent with previous DMRG studies~\cite{doi:10.1126/science.aal5304,PhysRevB.102.041106}, whereas electron doping yields an ordinary $d$-wave pattern. With increasing electron doping the correlations first become more extended, then progressively depart from $d$-wave symmetry.
Figure~\ref{fig:real_space_doping_dep} (on the right) shows the doping evolution of the static correlations $S_{\alpha,\alpha'}(\mathbf{r},\omega=0)$ at fixed temperature. 
For the same parameters, the static correlations are noticeably more short-ranged than their $\tau=\beta/2$ counterparts, reflecting that the $\omega=0$ correlation  places comparatively greater weight on higher-energy excitations.

\noindent\underline{Interaction dependence} \label{sec:suppinteractiondep}

\begin{figure}[tbp]
    \centering
    \includegraphics[width=\linewidth]{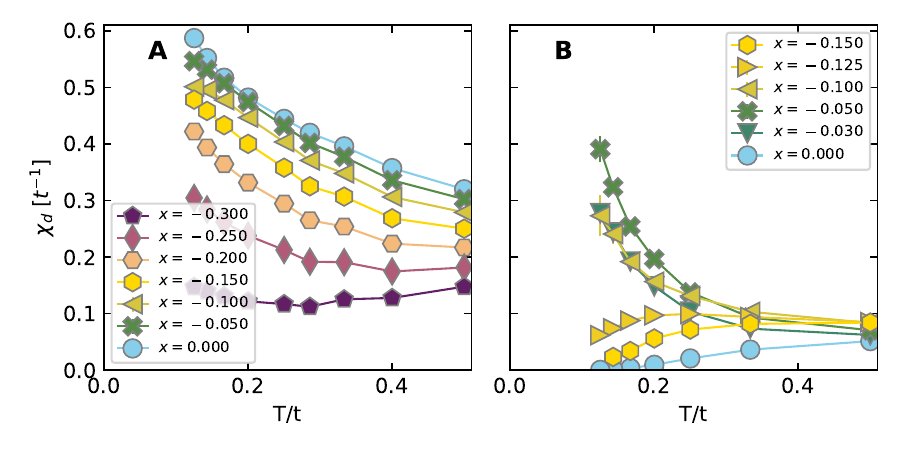}
    \caption{ Temperature dependence of $\chi_d$ for non-interacting vs interacting cases. (A) $U/t=0$, size $8\times 8$. (B) $U/t=6$, size $4\times 4$. Both: $t'/t=-0.25$. The interacting case uses a smaller size to access lower temperatures due to the fermion-sign problem.
}
    \label{fig:non-U}
\end{figure}

\begin{figure}[tbp]
    \centering
    \includegraphics[width=\linewidth]{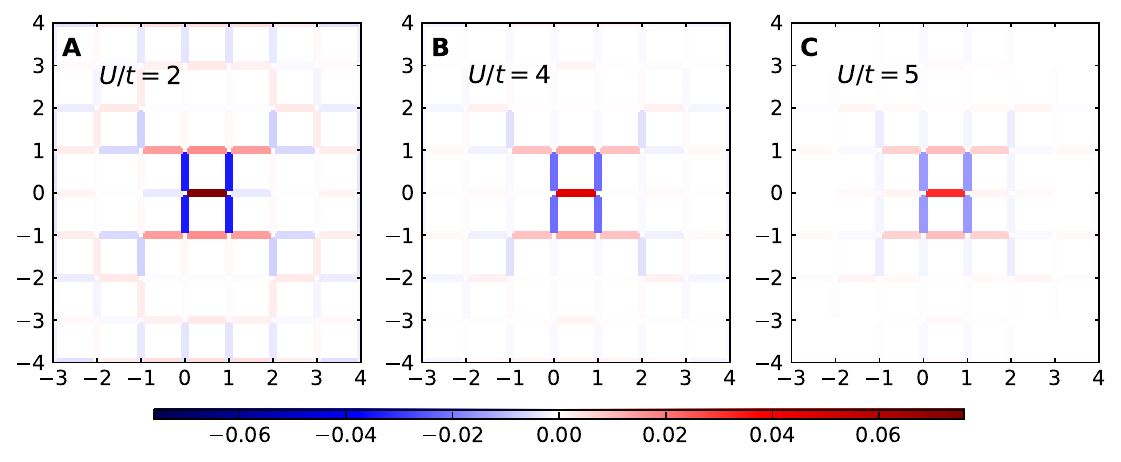}
    \caption{ Real-space pattern of the singlet-pair correlator $\beta S_{\alpha,\alpha'}(\mathbf{r},\tau=\beta/2)$ at electron doping $x=-5\%$ for various interaction strengths $U$.
    Temperature $T/t=0.2$; lattice size: $8\times 8$. We take the horizontal bond connecting $(0,0)$ and $(1,0)$ as the reference.
}
    \label{fig:realspace_Udep}
\end{figure}

In this section we contrast weak- and strong-coupling behavior of $\chi_d$ and $S_{\alpha,\alpha'}(\mathbf{r},\tau=\beta/2)$ in the electron-doped regime.
In Fig~\ref{fig:non-U}, We compare $U/t=0$ with $U/t=6$ (the interacting case uses a smaller lattice to access lower $T$ due to the fermion-sign problem). For $U/t=0$, $\chi_d$ decreases monotonically upon doping away from half filling and shows only weak temperature dependence across dopings.
By contrast, at $U/t=6$, the doping dependence is nonmonotonic, and within a doping window $\chi_d$ increases rapidly upon cooling, indicating an interaction-driven superconducting tendency absent at weak coupling.
Although at the accessible temperatures $\chi_d$ ($U/t=6$) remains smaller in magnitude than $\chi_d$	($U/t=0$), the contrasting temperature scaling suggests that at sufficiently low $T$ the interacting case will overtake the non-interacting one.
Consistent trends appear in the real-space correlator $S_{\alpha,\alpha'}(\mathbf{r},\tau=\beta/2)$, which acquires $d$-wave symmetry with stronger $U$ (Fig.~\ref{fig:realspace_Udep}).
Taken together, these observations indicate that the $d$-wave superconducting tendency is distinct from a weak-coupling scenario.

\noindent\underline{Finite size analysis} \label{sec:finitsize}

\begin{figure}[tbp]
    \centering
    \includegraphics[width=\linewidth]{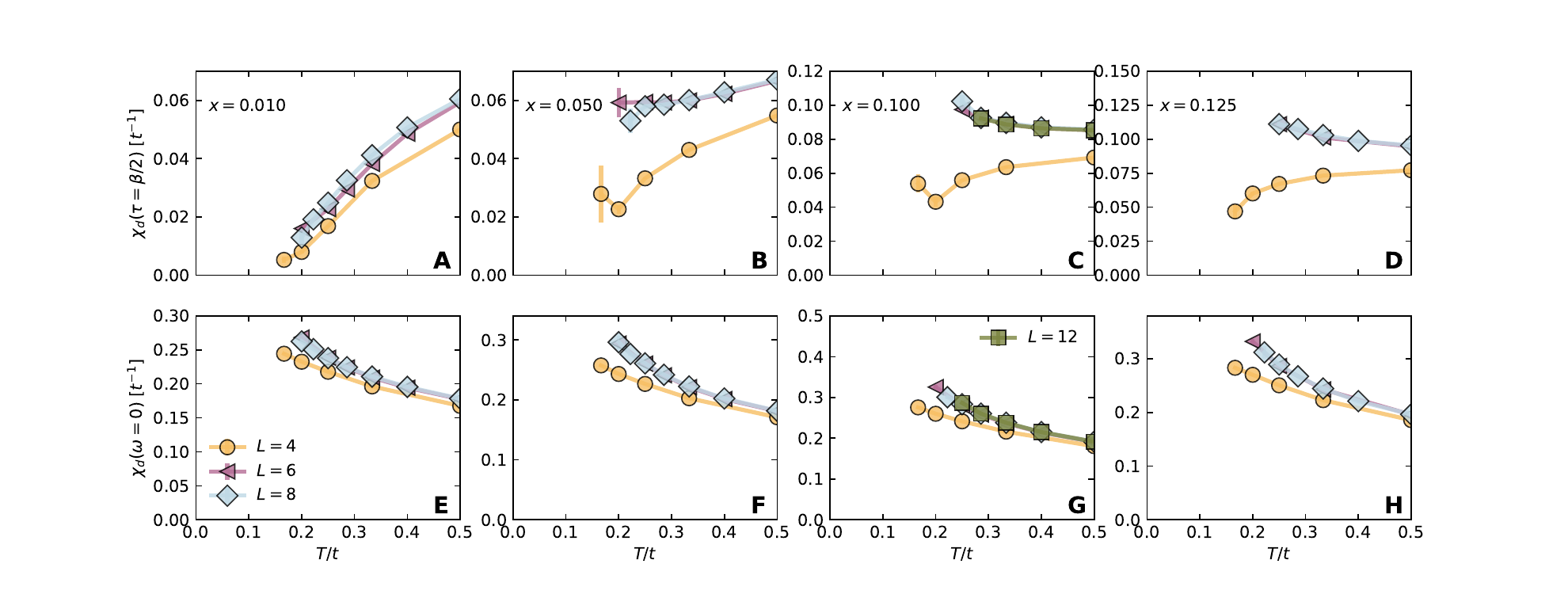}
    \caption{Pair-field susceptibility $\chi_d(\tau=\beta/2)$ (A-D) and $\chi_d(\omega=0)$ (E-H) on square lattices of size $L\times L$ for various hole-doping values. Parameters: $U/t=6$, $t'/t=-0.25$.
}
    \label{fig:size_dep_hole}
\end{figure}

\begin{figure}[tbp]
    \centering
    \includegraphics[width=\linewidth]{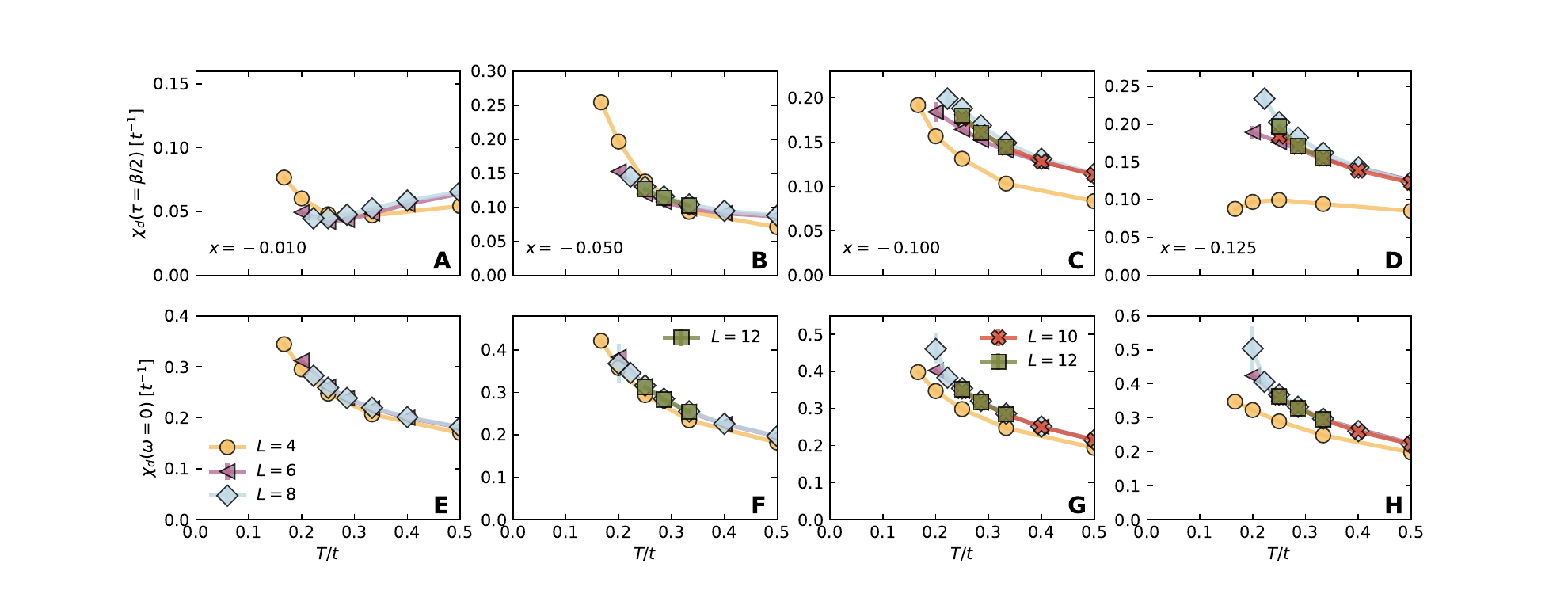}
    \caption{ Pair-field susceptibility $\chi_d(\tau=\beta/2)$ (A-D) and $\chi_d(\omega=0)$ (E-H) on square lattices of size $L\times L$ for various electron-doping values. Parameters: $U/t=6$, $t'/t=-0.25$.
}
    \label{fig:size_dep_elec}
\end{figure}

Figures~\ref{fig:size_dep_hole} and \ref{fig:size_dep_elec} show the size dependence of $\chi_d$ in hole and electron doping, respectively.
On the hole-doped side, finite-size effects are minimal for $L \times L$  lattices with $L \geq 6$, consistent with the absence of extended pairing correlations at the accessible temperatures.
On the electron-doped side, finite-size effects are more pronounced \cite{wang2025probing}, yet remain modest for $L \geq 6$ except when the doping exceeds $|x|\gtrsim 10\%$, where the correlations begin to approach the lattice boundaries.

We also find that the $\omega=0$ susceptibility exhibits weaker finite-size effects than its $\tau=\beta/2$ counterpart, consistent with its greater weighting of local/high-energy contributions.

\end{document}